\documentstyle[12pt,fleqn]{article}
\newcommand{\preprint}[1]{\hfill{\sl preprint - #1}\par\bigskip\par\rm}
\def\titolo{\par\bigskip\begin{center}\bf\LARGE}
\def\endtitolo{\end{center}\par\bigskip\par\rm\normalsize}
\def\instit{\begin{center}\large}
\def\endinstit{\end{center}\rm\normalsize}
\def\references{\begin{thebibliography}{10}}
\def\endreferences{\end{thebibliography}\end{document}}
\newcommand{\dip}{\smallskip Dipartimento di Fisica,
Universit\`a di Trento and Istituto Nazionale di Fisica Nucleare,
Gruppo Collegato di Trento,\\
Via Sommarive 14  I-38050 Povo (TN)
Italy}
\newcommand{\hostdip}{\smallskip Institut f\"{u}r Theoretische Physik der
Universit\"{a}t Innsbruck,
Victor-Franz-Hess-Haus,\\ Technikerstra{\ss}e 25/2 A-6020 Innsbruck Austria}

\newcommand{\btit}{\begin{titolo}}
\newcommand{\etit}{\end{titolo}}

\renewcommand{\author}[1]{\begin{center}\Large #1\end{center}}
\newcommand{\hostauthor}[1]{\begin{center}\Large #1\end{center}}
\renewcommand{\date}[1]{\par\bigskip\par\sl\hfill #1\par\medskip\par\rm}
\newcommand{\email}[1]{e-mail: \sl #1@alpha.science.unitn.it}
\newcommand{\hostemail}[1]{e-mail: \sl stefan.steidl@uibk.ac.at}
\newcommand{\femail}[1]{\footnote{\email{#1}}}
\newcommand{\hostfemail}[1]{\footnote{\hostemail{#1}}}
\newcommand{\pacs}[1]{\smallskip\noindent{\sl PACS number(s):
\hspace{0.3cm}#1}\par\bigskip\rm}
\newcommand{\babs}{\hrule\par\begin{description}\item{Abstract: }\it}
\newcommand{\eabs}{\par\end{description}\hrule\par\medskip\rm}
\newcommand{\ack}[1]{\par\section*{Acknowledgements} #1}
\renewcommand{\vec}[1]{{\bf #1}}       
\newcommand{\nn}{\nonumber}            
\newcommand{\beq}{\begin{eqnarray}}    
\newcommand{\eeq}{\end{eqnarray}}      
\newcommand{\beqn}{\begin{eqnarray}}   
\newcommand{\eeqn}{\end{eqnarray}}     
\newcommand{\R}{\mbox{$I\!\!R$}}                 
\newcommand{\Z}{\mbox{$Z\!\!\!Z$}}               

\newcommand{\be}{\beta}

\newcommand{\ep}{\varepsilon}

\newcommand{\om}{\omega}

\newcommand{\Om}{\Omega}

\begin{document}

\preprint{UTF 361 \\
gr-qc/9509036}

\btit
A Bisognano-Wichmann-like Theorem in a Certain Case of a 
{\em Non} Bifurcate Event Horizon related to an Extreme Reissner-Nordstr\"om
Black Hole
\etit

\author{Valter Moretti \femail{moretti}}
\dip

\hostauthor{Stefan Steidl \hostfemail{steidl}}
\hostdip

\date{July-August-September 1995}

\babs \\ 
Thermal Wightman functions of a massless 
scalar field
are studied within the framework of a  ``near horizon'' static 
background model
of an extremal R-N black hole. This model is  built up by using 
global Carter-like coordinates over an infinite set of
 Bertotti-Robinson submanifolds glued together. 
The  analytical extendibility beyond the horizon 
 is imposed as constraints  
on (thermal) 
Wightman's functions defined on a Bertotti-Robinson sub manifold. 
It turns out that only the Bertotti-Robinson
vacuum state, i.e. $T=0$, satisfies the above requirement.
Furthermore the extension of this state onto the whole manifold is
proved to coincide exactly with the vacuum state in the global 
Carter-like coordinates.   
Hence a theorem similar to
Bisognano-Wichmann theorem for the Minkowski space-time in terms of 
Wightman functions holds with vanishing ``Unruh-Rindler temperature''.
 Furtermore, the Carter-like vacuum restricted to a Bertotti-Robinson
region, resulting a pure state there, has vanishing entropy despite of the
presence of event horizons.
Some comments  on the real extremal R-N black hole are given.   
\eabs 


\newpage

\section*{Introduction}

In a space-time with {\em non empty} intersection event horizons,
i.e. Rindler-Schwarzschild-like  space-time, several methods for determining
 the possible equilibrium state of a
scalar field propagating therein exist.  They select 
one special temperature only, the Rindler-Unruh-Hawking temperature.
These theorems use the KMS condition \cite{kms} and the 
 Haag Narnhofer Stein principle
(i.e., the  "Haag scaling prescription") \cite{HNS,FH}  or demand   
a  stationary and Hadamard behaviour of the Wightman functions
\cite{kaywald}.\\
These theorems can not be employed in the case of an extremal
Reissner-Nordstr\"{o}m black hole due to the appearence of 
a  null  surface gravity as well as a
{\em non-bifurcate event horizons}. In fact the future event horizon
and the past event horizon do not intersect there.
However, Anderson, Hiscock and  Loranz \cite{AHL} proved that only the
Reissner-Nordstr\"{o}m vacuum state has a regular stress-tensor on the
horizon
and thus only this state is a  possible
equilibrium state in the framework
of semiclassical quantum gravity.
Finally, in  recent works \cite{moretti, moretti2}, Moretti shows that the 
Haag Narnhofer and Stein principle for the behaviour of 
Wightman's function on the horizon of a black hole results to be unable to 
determine the really admissible thermal quantum states
in the case of an extremal R-N black hole, but a further
development of
the previous principle, the  Hessling principle \footnote{Roughly 
speaking, these two principles correspond respectively to a weaker and a
 stronger version of a Quantum 
Einstein's Equivalence Principle. They  require a weaker and a 
stronger ``Minkowskian behaviour''
 of two point Wightman functions in the limit of vanishing geodesical distance
between the arguments.}\cite{hessling,moretti,moretti2} determines only 
the Reissner-Nordstr\"om quantum vacuum as physically admissible (i.e. $T=0$).
Similar result, but using very different analysis,
 appeared in \cite{vanzo}.
These facts seem to improve the topological 
 result obtained by the method of the 
elimination of conical singularities from the Euclidean,
 time extended manifold, which accepts any value of the temperature for 
a R-N black hole\cite{HHR,GM}.\\
Almost all the previously mentioned papers deal with  quantum field states  
at least defined in a certain space-time region (boundary included)  bounded
by event horizons, e.g. 
 the external
region of a black hole. \\
On the other hand, it is obvious that the classical field is not blocked by 
the horizons and thus it seems to be necessary to demand the
 existence
of  global extensions  of physically sensible 
quantum field states.  This request result to be 
 satisfyed by the Minkowski vacuum in the
Rindler wedge theory and by the Hartle-Hawking
 state in the Schwarzschild 
black hole theory. \\
As well-known,
 the extremal Reissner-Nordstr\"om manifold can be maximally extended into 
Carter's manifold and thus it seems to be  interesting to study 
quantum field states defined on the whole Carter manifold (if they exist).\\
 In this paper, we shall study a ``near horizon'' model of
Carter's and Reissner-Nordstr\"om manifolds. 
Following an algebraic approach to quantum field theory and
 starting from KMS quantum states  initially defined  
in a Reissner-Nordstr\"om-like submanifold only,
we shall study
the existence of  {\em analytical} extensions beyond the horizons of 
their Wightman functions and thus in the whole Carter-like manifold.\\
In particular, as our first result,
 we shall prove the  possibility of an
 algebraic quantum
field formulation on our manifold despite of the fact that 
this is
 non globally hyperbolic.\\ 
Moreover, as our second result,
 we shall prove that only the approximated vacuum state
 corresponding
to the Reissner-Nordstr\"{o}m vacuum state (with zero temperature)
can be extended beyond the horizons.\\
Furthermore 
we shall see  that there exists a  relation between our model-manifold
endowed with Bertotti-Robinson sub-manifolds 
and Minkowski space-time endowed with the well-known couple of
Rindler wedges. These two 
structures act as ``toy models'' of 
two different kinds of black holes: the extremal R-N black hole and 
the eternal Schwarzschild black hole respectively.
Implementing this analogy, as our third result,
 we shall recover the equivalent of the 
the {\em Bisognano-Wichmann theorem} for the Minkowski space-time\cite{sewell}.
In this contest, the analog of the Minkowski vacuum
is the vacuum defined with respect to the global Carter-like
coordinates of our manifold.
The $\be=2\pi$-Rindler-KMS state corresponds to
the vacuum of the R-N-like coordinates.\\
Thus, an  important difference arises.  
The analog of the Rindler-Unruh temperature is now  {\em zero} 
and thus  {\em no} KMS prescription appears, but the {\em stationarity}
 of the state 
remains, i.e., the  functional dependency of only the difference of 
the temporal arguments.\\
In {\bf Section 1} we shall introduce the well-known  Carter representation 
for a maximally analytically extendible manifold for an extremal R-N black 
hole. Furthermore, we shall perform the necessary
approximations in order to deal with a
neighbourhood of the  horizon.\\
In {\bf Section 2}, using approximated Carter coordinates and the 
Bertotti-Robinson metric,  we shall construct a ``near horizon'' toy model
of Carter's manifold which results to be  non globally hyperbolic.
We shall prove that it is possible to define 
 a quantum field theory.   Finally, 
we shall study  the analytical extension 
of Wightman's functions  beyond the horizons proving also 
a Bisognano-Wichmann-like theorem in terms of Wightman functions.\\
 In {\bf Section 3}, we shall point out our conclusions and we shall
look at the real extremal  R-N black hole.


\section{Carter's Manifold and Approximations near its Horizons}

\input epsf
 
The general form of Reissner-Nordstr\"om black-hole metric is given by
\cite{hawking libro}
\beq
ds^{2}= -\left( 1-\frac{2M}{\bar{r}}+ \frac{Q^2}{\bar{r}^2}
\right)^{2} dt^{2} +
\frac{1}{ \left( 1-\frac{2M}{\bar{r}}+ \frac{Q^2}{\bar{r}^2}\right)^2}
 d\bar{r}^{2} +
\bar{r}^{2}\left( d\theta^{2}+\sin^{2}\theta d\varphi^{2}\right)
\nn \:,
\eeq 
where $M$ is the mass and $Q$ the charge of the black hole.
We are interested in {\em extremal} case $Q=M$. Thus we have
\beq
ds^{2}= -\left( 1-\frac{M}{\bar{r}}\right)^{2} dt^{2} +
\frac{1}{\left( 1-\frac{M}{\bar{r}}\right) ^{2}} d\bar{r}^{2} +
\bar{r}^{2}\left( d\theta^{2}+\sin^{2}\theta d\varphi^{2}\right)
\nn \:.
\eeq 
For sake of simplicity, we shall chose an extremal  black hole of
unit mass by a suitable choice of the units of measure.
In addition, we shall use the abbreviation $d\Omega_{2}
:= \left( d\theta^{2}+\sin^{2}\theta d\varphi^{2}\right)$.
Thus, our metric reads  
\beq
ds^{2}= -\left( 1-\frac{1}{\bar{r}}\right)^{2} dt^{2} +
\frac{1}{\left( 1-\frac{1}{\bar{r}}\right) ^{2}} d\bar{r}^{2} +
\bar{r}^{2}d\Omega_{2}\label{RN} \:.
\eeq 

As well-known, the above chart
does not cover the
whole manifold as $ds^{2}$ is singular at $\bar{r} = 1$. 
This inconvenience can be avoided, following the Schwarzschild case,
by  introducing  Kruskal-like
 coordinates\footnote{It must be stressed that in contrast to Schwarzschild 
coordinates in the interior of the event horizon, where, roughly speaking, 
the temporal and radial coordinates change roles (in the usual 
interpretation), the coordinates $\{\bar{r}, t\}$ in the R-N chart 
keep their meaning also for $\bar{r} < 1$.}, i.e., {\em Carter's coordinates}
\cite{carter}. These  define a {\em maximally 
analytically extended} manifold  obtained from the 
R-N manifold $(\bar{r}>1)$. 
To begin with, we introduce two functions $u(t,\bar{r})$ and $w(t,\bar{r})$ 
by
\begin{eqnarray}
u &=& r^{\ast} + t \:,\label{X}\\
w &=& r^{\ast} - t \:, \label{Y}
\end{eqnarray}
where $r^{\ast}$ is given by the  invertible
function of $\bar{r}>1$
\beq
r^{\ast}(\bar{r}) = \int
\frac{d\bar{r}}{\left( 1-\frac{1}{\bar{r}}\right) ^{2}} =
\bar{r}\frac{\bar{r}-2}{\bar{r}-1}+2\ln \left| \bar{r}-1\right|\label{ara} \:.
\eeq
Let us  introduce Carter's coordinates in the Reissner-Nordstr\"om manifold
$\{T, R,\theta ,\varphi\}$
\cite{carter}\footnote{ 
In this paper, in order to simplify calculations, 
we define Carter's coordinates changing  original Carter's 
definition by a trivial linear 
transformation, $T=T_C/2- 3\pi/4$, $R=R_C/2+\pi/4$. Where $R_C$ and $T_C$
are  Carter's coordinates as they appear in \cite{carter}. } through
 the equations
\begin{eqnarray}
u & = & - \cot \left( T + R \right)\label{Z}\:,\\
w & = & + \cot \left( T - R \right) \label{T}\:,
\end{eqnarray}
and thus
\begin{eqnarray}
2T & = & \cot^{-1}(w) - \cot^{-1}(u) \:,\\
2R & = &   -\cot^{-1}(w) -\cot^{-1}(u)\:,
\end{eqnarray}
where $T\in ] -\pi/2, +\pi/2[$ and $R\in ] 0, \pi[$.
The metric in Eq.(\ref{RN}) reads now
\beq
ds^{2} = Q\left( -dT^{2}+dR^{2}\right) + \bar{r}^{2}
d\Omega_{2} \label{RNC}\:,
\eeq
where
$Q$ is given by
\beq
Q =
\left( 1-\frac{1}{\bar{r}}\right)^{2}\csc^{2}\left( T+ R 
\right)
\csc^{2}\left( T - R \right) \label{q}\:.
\eeq
This form of metric can be  extend to a larger manifold 
 where $T$ ranges from $-\infty$ to $+\infty$ and $R$ ranges, 
from  $0$ to $2\pi$ (the angular variables having their customary
range).\\
A part of the complete manifold is represented in {\bf figure~1}. The initial
form of the metric (\ref{RN}) holds in each of the R-N regions, conversely,
the new form (\ref{RNC}) holds in the whole manifold.\\ 
Note that the right edges as well as the intersection points of
the horizons  of the infinite number of R-N zones
\beq
R=0 \:\:\:\: T=0,\: \pm \pi, \:\pm2 \pi,\: \pm3 \pi,\: ...
\eeq
are not in the manifold (the diagram is really a {\em Penrose's diagram}).
In fact, these points are infinitely far away from internal points of the  
manifold if the distance along time-like or space-like
geodesics is taken or  the affine parameter distance along
light-like geodesics is considered. From this property 
it also follows that it is not possible to  (analytically) extend  
the manifold any further.\\
The {\em time-like} 
irremovable singularity is represented by all the points which have 
$R=0$ and $T$ taking values different from $k\pi$, $k\in \Z$.
The open R-N regions, i.e., the 
R-N regions without boundary and event-horizons, are 
globally hyperbolic, the ``lines'' $T= k \pi$ being  Cauchy-surfaces.\\
Returning to Eq.(\ref{q}), we  observe that,
in order to define  $Q$  on the whole manifold,
one has to  analytically extend and then invert (in the variable $\bar{r}$)
\beq
-\cot\left( T + R \right) +
\cot\left( T - R \right) =
2 r^{\ast}(\bar{r}) = 
2\bar{r}\frac{\bar{r}-2}{\bar{r}-1}+4\ln \left| \bar{r}-1\right|
\label{r estesa} \:,
\eeq
Then, by means of Eq.~(\ref{r estesa}),
it is possible to restore the same form of the metric of 
Eq.~(\ref{RN}) also {\em outside} of the Reissner-Nordstr\"{o}m
region, excluding  the horizons. 
In fact, once one uses Carter's coordinates, one may redefine the $r^{\ast}$
variable outside of the R-N region (towards the future horizon for example) 
by Eq.~(\ref{r estesa}) and the $\bar{r}$
variable by means of Eq.(\ref{ara}) as $\bar{r}<1$. Finally the $t$ variable
is restored by a trivial use of Eq.s (\ref{X}), (\ref{Y}), (\ref{Z}), 
(\ref{T}). 
Note that, in this way, one can also construct a 
{\em time-like Killing vector} on the 
whole manifold (with the exception of  the horizons) 
simply by considering the tangent vector to the $t$ coordinate. 

\begin{center}
\leavevmode
\epsfbox{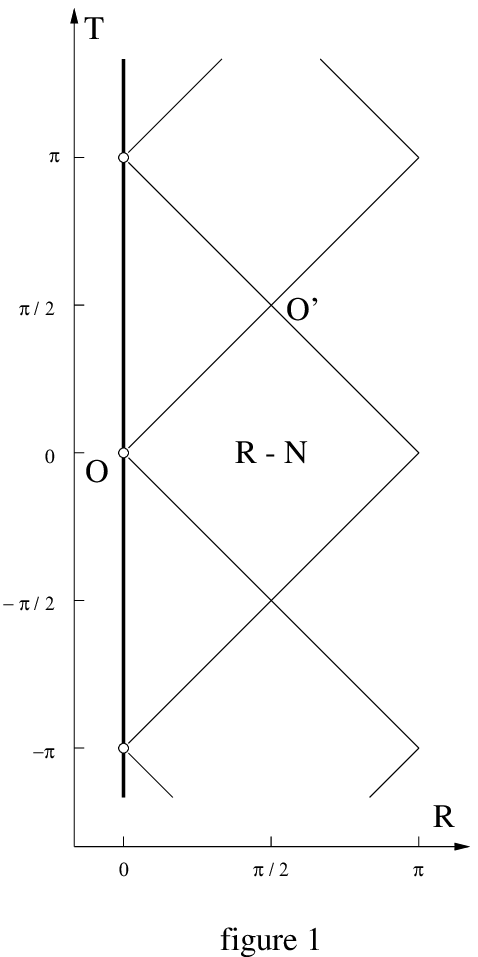}
\end{center}
\bigskip

To conclude, we observe that  Carter's manifold is not the only
manifold which one can build up starting from R-N's manifold. For
example it is possible to identify two local charts of Carter's chain
and thus obtain a new manifold containing a  finite number of R-N zones.
However this kind of extension trivially 
violates the {\em weaker causality condition}\cite{wald},
hence it is not clear whether a quantum field theory can be defined
there\footnote{However, Kay et al. investigated the possibilities of
a QFT in similar backgrounds \cite{kay,higuchi} recently. }.\\

Now we shall consider an  approximated  metric near the horizons.
 In the shaded 
region near the horizons represented in {\bf figure 2}, the metric
(\ref{RNC}) can be approximated by a {\em static} metric 
\beq
ds^2 \sim ds^2_0 := \frac{4}{\sin^2 2R} (-dT^2+dR^2) + d\Omega_2 \label{dso}
\:. 
\eeq
The vector $\partial_T$ defines an approximated {\em time-like Killing vector}
near the horizons.\\
In the same region, but considering R-N coordinates, $ds^2$ can be 
approximated by the {\em Bertotti-Robinson} metric \cite{BR}, as well-known
 \cite{referee}. Thus we have
\beq
ds^2 \sim ds^2_{BR}:= \frac{-dt^2+dr^2+d\Omega_2}{r^2} \label{aBR}\:,
\eeq
where
\begin{eqnarray}
r&:=& -r^\ast \:\:\:\: \mbox{if} \:\:\:\: R>T \:\:\:\: \mbox{or} \\
r&:=& +r^\ast \:\:\:\: \mbox{if} \:\:\:\: R<T \:.
\end{eqnarray}

\begin{center}
\leavevmode
\epsfbox{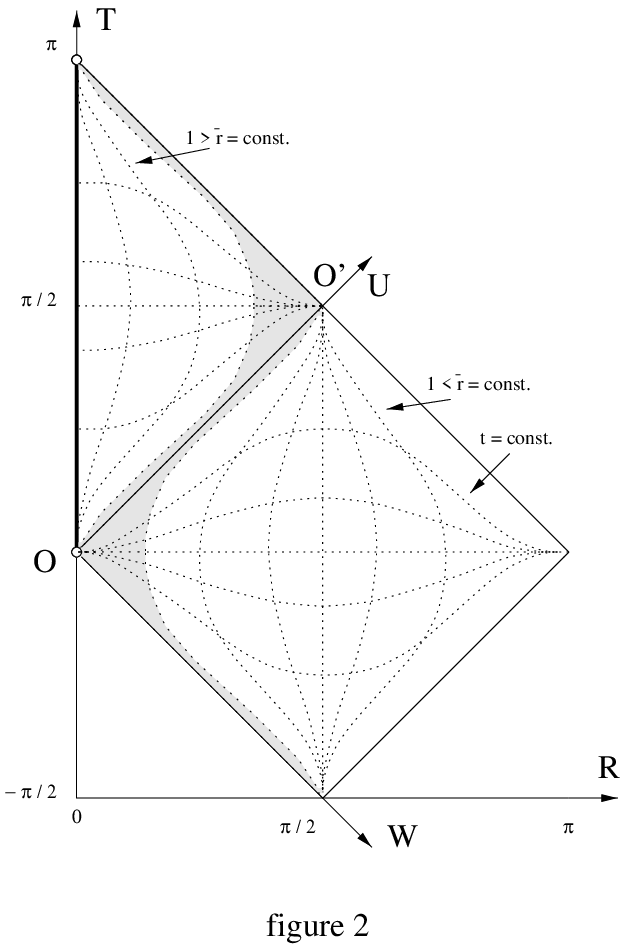}
\end{center}
\bigskip

Finally, in the considered region,
 the transformation law between $r,t$ and $R,T$
is 
\begin{eqnarray}
2 r &\sim& | \cot (R-T) +\cot(R+T) | = \frac{2\sin 2R}{|\cos 2T\: -\cos2R |}
  \:,
\label{ar}\\
2 t &\sim& \:\:\cot (R-T) -\cot(R+T) \:\: =\frac{2\sin 2T}{\cos 2T\: -\cos2R }
\:,
\label{at}
\end{eqnarray}
All the previous approximations are carefully 
examined in {\bf Appendix A} (see also \cite{referee}).

\section{Our Toy Model and its Thermal Wightman Functions}
\input epsf

In this section we consider the Wightman functions of a massless scalar
field 
obtained by quantizing it
in  a new manifold, built up by using the Bertotti-Robinson metric only.
Obviously  we suppose these Wightman functions 
approximate to the ``true'' Wightman functions of the 
Reissner-Nordstr\"{o}m metric
near the horizon (inside the R-N zone), where the two metrics are
undistinguishable.
A similar  hypothesis  was also used to calculate the
renormalized stress tensor. Then, an  independent check proved that this 
approximation was correct
for that purpose at least \cite{AHL}. Furthermore, 
a similar assumption was used to obtain the Hawking temperature 
in the case of
a Schwarzschild black hole \cite{HNS} 
and the result was proved to be correct.\\
In {\bf Appendix B} we return on this assumptions with some general 
mathematical comments.\\

 In order to have a mathematically well defined background of our field
theory, we shall build up a complete manifold by gluing together an infinite 
number of Bertotti-Robinson charts as pointed out in {\bf figure~3}.
$T$ and $R$ are {\em global coordinates} of this manifold. These are
connected to Bertotti-Robinson variables $t$ and $r$, in every B-R region,
by the following equations (see Eq.s (17) and (18))
\begin{eqnarray}
2 r &=& | \cot (R-T) +\cot(R+T) | = \frac{2\sin 2R}{|\cos 2T\: -\cos2R |}  \:,
\label{r}\\
2 t &=& \:\:\cot (R-T) -\cot(R+T) \:\: =\frac{2\sin 2T}{\cos 2T\: -\cos2R }\:,
\label{t}
\end{eqnarray}
where $R\in ] 0, \pi/2 [$ and $T\in \R$.
It can be easy shown that, considering the form of the metric and
the relations between  $r,t$ and $R,T$ 
near every horizon, one find the same equations as in the previous
section. In this sense our manifold is a  toy model of Carter's
manifold. The global form of the metric (which is regular on the 
horizons) is the  static metric of Eq.~(\ref{dszero})
\beq
ds^{2} = \frac{4}{\sin^{2}2 R}(-dT^{2} + dR^{2} +
\sin^{2}2R \:d\Om_{2})\:,\label{quasi einstein}
\eeq
The above metric is
 conformal to the metric of the
{\em Einstein static universe}\footnote{Usually, the metric of Einstein's
static universe is written in terms of $R':=2R$ and $T':=2T$
so that the global factor $4$ disappears.} by the factor $1/\sin^{2}2R$.\\
Let us  look at our manifold as it is represented in {\bf figure~3}.\\
The edges look like singularities of the
metric, but it is not the case. In fact, the intersection points
of the horizons ($r=+\infty$) are not in the manifold because each
geodesic  reaching them from an inner point
spans an infinite Riemannian length (or an infinite affine
parameter gap if it is a  null geodesic).
Similarly, by  calculating also the geodesics which reach
$r=0$,
it is possible to prove the same property
for all the remaining points on the manifold's edges.
Hence the edges are not in of the manifold
and thus it is  not possible to extend
the manifold any further.\\

\begin{center}
\leavevmode
\epsfbox{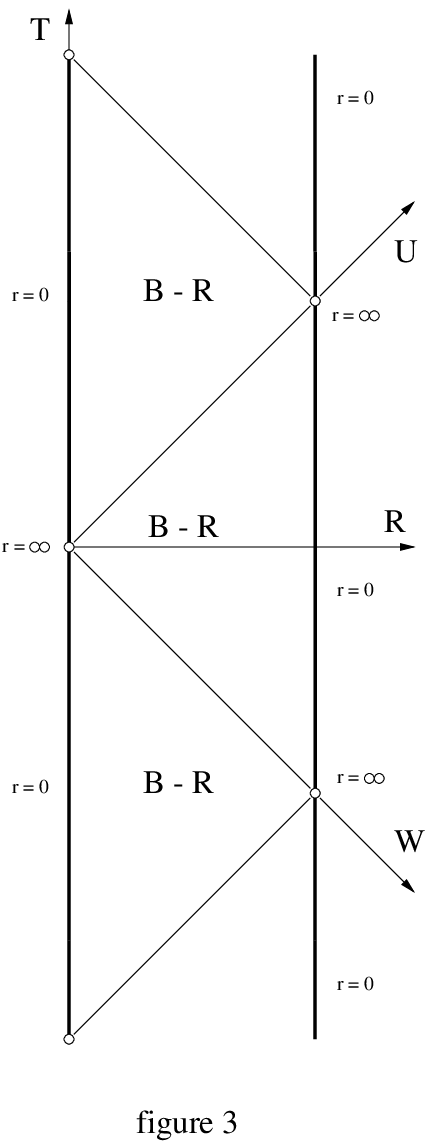}
\end{center}
\bigskip

Finally we stress that passing to the Euclidean time $t_{E}= it$
and using $r,t_{E}$ coordinates, {\em no conical singularity}
arises for any choice of the Euclidean time period $\be$\footnote{The point
$r=0$ which (which is a conical singularity of the Euclidean manifold in the
Rindler or Schwarzschild cases) does not belong to the manifold now.}. Hence,
following \cite{HHR} and \cite{GM}, we should accept all the values of
the temperature of the KMS states defined inside of a B-R zone.\\

It is possible to compare  Carter's 
manifold to Kruskal manifold in the 
following sense.\\
  Let us consider Kruskal manifold.
There, the metric looks like that of a Rindler space if Schwarzschild's
coordinates are used, 
or,
 that of a Minkowski space if  Kruskal's coordinates are adopted. Also, the 
transformation  laws between these two coordinate systems locally resemble the 
corresponding transformation laws in the Minkowski manifold.
Furthermore, near the horizon, the Kruskal time defines  an approximated 
time-like Killing vector which becomes a global time-like Killing vector
in the Minkowski space-time.\\
Considering Carter's manifold, the same features will arise.
 We have to consider Carter's coordinates as Kruskal's coordinates,
our global Carter-like model as a Minkowski's manifold and 
Bertotti-Robinson manifold as a Rindler wedge.
Then, near the horizons, 
Carter's metric looks like that of Bertotti-Robinson if we use
Reissner-Nordstr\"{o}m coordinates, or the metric conformal to the Einstein
static metric if we used Carter's coordinates and so on. In particular,
the approximated time-like Killing vector near the horizon in Carter's
manifold becomes
an exact time-like  Killing vector in our Carter-like global manifold.\\ 
Roughly speaking, the quantum field theory in a Rindler wedge
on the background of  a Minkowski space-time appears as a simplified quantum
field theory in a Schwarzschild space-time on the background of a  
Kruskal manifold.  The coincidence of the Unruh-Rindler state with
the Minkowski vacuum (Bisognano-Wichmann theorem)
  appears as a simplified version of the
coincidence of $\be= 4\pi$-Schwarzschild-KMS state \footnote{Using opportune
units of measure.}
with the Hartle-Hawking 
state. It is reasonable to expect a similar situation for a quantum field 
theory on the Carter manifold. \\
In the following we want
to implement a part of this  idea proving, as our  third result,
 the equivalent of
the Bisognano-Wichmann theorem  in our Carter-like manifold. \\
Like  Carter's manifold, our Carter-like manifold is 
  not globally hyperbolic
\cite{wald, hawking libro} because near the edge $r=0$
it is possible to find a pair of points  $p$, $q$ with $J^{+}(p) \cap
J^{-}(q)$ not closed and thus  not compact; furthermore,
 differently
 from Carter's manifold, all the ``patch manifolds'' (B-R zones)
are  non globally hyperbolic. We shall prove,
that inside any of
these regions as well as in the whole Carter-like manifold, 
a ``quasi-standard'' quasifree scalar field theory can be defined.

\subsection{Possibility of a QFT}

In order to point out the possibility of a QFT on our manifolds,
we shall follow the {\em algebraic approach}
 used in \cite{kaywald} based on 
{\em Weyl algebra}.\\ 
First we consider the B-R submanifolds.
 From  now
on, we shall understand $x^{0},x^{1},x^{2},x^{3}$ (posing also $t:=x^{0}$,
$\vec{x}\equiv (x^{1},x^{2},x^{3})$ and
$r:=\mid \vec{x}\mid$)  as Minkowski coordinates in Minkowski space
as well as  Bertotti-Robinson coordinates in the Bertotti-Robinson space.\\
Let us start by considering that  a  very
simple connection between the solutions of the massless Klein-Gordon equation
in Minkowski space and in the Bertotti-Robinson space exists.
If $\phi(\vec{x},t)_{M}$ indicates a generic $C^{\infty}$
solution with compact spatial support
of the massless Minkowskian K-G equation, then
\beq
\phi(\vec{x},t)_{BR} = r \phi(\vec{x},t)_{M}\:, \label{identita'}
\eeq
where $r > 0 $, $t \in \R$ and $\phi(\vec{x},t)_{BR}$ is a
solution of the massless Bertotti-Robinson K-G equation of the same order
of smoothness but without compact spatial support in general.
In order to build up the Weyl algebra
\cite{haag libro,fulling,kaywald},
we have to consider the following 
 bilinear symplectic form or indefinite scalar
product 
\beq
\sigma(\phi_{1},\phi_{2}) :=\int_{t=\mbox{const.}}\phi_{1}
{\stackrel{\leftrightarrow}{\nabla}}_{\mu}\phi_{2}\:
n^{\mu} \sqrt{h}\:\: dx_{1}
dx_{2} dx_{3}\:, \label{simplettica}
\eeq
where $n$ is the normal (and normalized) vector to the {\em Cauchy}
 surface $t=$constant and
$h$ is the determinant of the induced metric on this surface.\\
In the case of Minkowski space the above surfaces are Cauchy surfaces, but
this is not true in the case of Bertotti-Robinson space and thus we can not
deal with the standard theory.
However, if we decide to restrict
the possible vector space $S$ of solutions of the K-G equation by considering
only the scalar fields
on the left hand side of Eq.~(\ref{identita'})\footnote{Roughly 
speaking, this is similar  to a 
boundary condition requirement on the fields
solutions of the K-G equation in the B-R manifold.}, we shall trivially
find the following identity
\beq
\sigma(\phi_{BR1},\phi_{BR2})_{BR}=\sigma(\phi_{M1},\phi_{M2})_{M}\:.
\eeq
Following the notations of \cite{kaywald}, we can formally define the ``field
operator'' $\hat{\phi}_{BR} $ on B-R manifold by posing
\beq
\sigma(\hat{\phi}_{BR},\phi_{BR})_{BR}=\sigma(\hat{\phi}_{M},
\phi_{M})_{M}\:,
\eeq 
where $\phi_{BR} \in S$.
Starting from the set $S$ and using the previous identities one can construct
the usual theory of quantum fields for quasifree states in the
algebraic
approach based on the Weyl algebra (\cite{kaywald})
 on the Bertotti-Robinson background,
too.
In particular, (quasi-free) states can be built, using the (quasi-free)
states of Minkowski theory as follows
\beq
\lambda(\phi_{BR1}...\phi_{BRn})_{BR} :=
\lambda(\phi_{M1}...\phi_{Mn})_{M} \:, \label{F}
\eeq
where $\lambda_{BR}$ ( $\lambda_{M}$) denotes a generic $n$-point function 
evaluated on the K-G
solutions $\phi_{BRk}$  ($\phi_{Mk}$) and these fields are
related to each other by Eq.~(\ref{identita'}).\\
In terms of integral kernels the above identity reads as
\beq
\lambda(x_1,x_2,...x_n)_{BR} :=
r_1 r_2...r_n\:\:\lambda(x_1,x_2,...x_n)_{M} \:, \label{F'}
\eeq 
We note that this formulation of
quantum field theory in B-R space-time agrees  with 
Kay's general formulation for generally {\em non} globally hyperbolic 
manifold based on {\em F-locality} \cite{kay,higuchi}.
This follows  using
test functions with compact support {\em inside of the B-R manifold},
 re-formulating the theory in terms of  
a ``four-smeared'' field theory and using the ``advanced minus retarded''
Green function induced by antisymmetryzing Eq.(\ref{F'}) for $n=2$. 
\footnote{Successively, 
one could use the Fewster-Higuchi theorem \cite{higuchi}, for example.}\\
The thermal Wightman
functions relative to the vacuum state arising by canonical quantization
in Bertotti-Robinson coordinates obviously satisfy
\beq
W_{\be}^{\pm}(x,x{'})_{BR} = rr^{'}\:W_{\be}^{\pm}(x,x{'})_{M}\:,
\eeq
which follows directly from Eq.~(\ref{F'}).\\ 
Summing over the normal modes of the Minkowskian K-G equation and then
thermalizing by employing
the well-known {\em sum over images method} \cite{fr,mahan,kapusta,mova}
we easily
obtain the thermal Wightman functions
(really distributions) for a massless
scalar field in the B-R background.
They read, dropping the index 'BR'
\beq
W^{\pm}_{\be}=   \frac{\mid \vec{x}\mid  \mid \vec{x^{'}}\mid}{4\pi^2}
\frac{\pi\left\{\coth\frac{\pi}{\be}
(|\vec{x}-\vec{x}^{'}|+t-t^{'}\mp i\ep)+\coth
\frac{\pi}{\be}(|\vec{x}-\vec{x}^{'}|-t+t^{'}\pm i\ep)\right\} }{2\be
|\vec{x}-\vec{x}^{'}|} \label{wightman distribuzione}
\eeq
Or, for $T= 1/\be =0$
\beq
W^{\pm}(x,x^{'})= \frac{\mid \vec{x}\mid  \mid \vec{x^{'}}\mid}{4\pi^2}
\frac{1}{ \mid \vec{x} - \vec{x^{'}}\mid^{2} -
\left(t - t^{'}\mp i\ep\right)^{2} }
\label{wightman zero distribuzione} \:.
\eeq
We observe that, in the sense of usual limit of functions
 but also using a weak limit interpretation,
one obtains the second Wightman function  as  
$\beta \rightarrow +\infty$ in the first one.\\
Note that the forms (\ref{wightman distribuzione}) and
(\ref{wightman zero distribuzione}) for the
Wightman functions hold in the interior of every B-R zone of the 
complete manifold only.\\
In order to prove the possibility of a QFT
on the whole Carter-like manifold we shall use the {\em 
Dowker-Schofield scaling property} \cite{dowkerschofield},
generalizing the previous proof. 
 Remind that the 
static Einstein's universe 
is globally hyperbolic and thus a standard algebraic QFT can be defined
there.
Following Dowker and Schofield \cite{dowkerschofield},
 let us  suppose to have two {\em static} metrics
which are conformally related
\beq
ds^{2}=g_{00}(\vec{x})(dx^{0})^{2}+g_{ij}(\vec{x})dx^{i}dx^{j} \label{cuno}
\eeq
and
\beq
ds^{'2}=g^{'}_{00}(\vec{x})(dx^{0})^{2}+g^{'}_{ij}(\vec{x})dx^{i}dx^{j}\:,
\label{cdue}
\eeq
where
\beq
g^{'}_{\mu\nu}=\lambda^{2}(\vec{x}) g_{\mu\nu}
\eeq
and let us  consider the solutions of the
respective Klein-Gordon-like equations
\beq
\left(\Box +\xi R + m^{2}\right) \phi(x) =0  \label{KG1}
\eeq
and
\beq
\left(\Box^{'} +\xi R^{'} +\left(\xi-\frac{1}{6}\right)\Box^{'}
\left(\lambda^{-2}\right)+  m^{2}\lambda^{-2}\right)
\phi^{'}(x)=0\:,  \label{KG2}
\eeq
where $R$ is the scalar curvature. Then,
the solutions  of the Klein-Gordon equations above are connected to 
each other by the
Dowker-Schofield scaling
property \cite{dowkerschofield} 
\beq
\phi(x)= \lambda(\vec{x}) \phi^{'}(x)  \:, \label{scaling}
\eeq
In our case, we consider $ds^{'2}$ as 
the metric of Einstein's static universe and
$ds^{2}$ as the metric
of Eq.~(\ref{quasi einstein}). Thus $\lambda^{2}=\sin^{2}2R$.\\
The fields propagating in the whole Carter-like manifold
 satisfy the Klein-Gordon
equation (\ref{KG1}) with $m=0$ and $R=0$. In fact the metric of
Eq.~(\ref{quasi einstein})
has vanishing scalar curvature.
We are free to choose $\xi=1/6$ ({\em conformal coupling}).
 Thus, if $\phi_{ESU}$ is a solution of the massless,
{\em conformally coupled} Klein-Gordon equation in Einstein's Static Universe
($\equiv$'$ESU$'), the field
\beq
\phi(\vec{R},T)_{CEU} := \sin{2R}\:\:\phi(\vec{R},T)_{ESU}\:, \label{esu}
\eeq
will satisfy the massless,
{\em minimally coupled} Klein-Gordon equation with the metric
(\ref{quasi einstein}) ('$CEU$'$\equiv$ 'Conformal to Einstein's Universe').
Due to Eq.(\ref{esu}), one finds also
\beq
\sigma(\phi_{CEU1},\phi_{CEU2})_{CEU} =\sigma(\phi_{ESU1},\phi_{ESU2})_{ESU}\:.
\eeq
Like in the previous case, it is possible to find an algebraic (quasi-)standard
field theory for the metric (\ref{quasi einstein})
by starting from  the vector space $S'$ of K-G solutions defined by
Eq.~(\ref{esu}), while the right hand side covers the vector space of
(conformally coupled) K-G solutions in the Einstein's static universe
of the class $C^{\infty}$\footnote{It is not necessary to demand
a spatial compact support because the spatial section of Einstein's
static universe is compact as well known, being homeomorfic to $S^{3}$.
Moreover, the observation above on Kay's
F-locality remains valid also in this case.}.\\
Finally, the Wightman functions satisfy  ($T=1/\be \geq 0$)
\beq
W_{\be}^{\pm}(X,X^{'})_{CEU} =
\sin^{2}2R \: \sin^{2}2R^{'} \:\: W_{\be}^{\pm}(X,X^{'})_{ESU}\:.
\label{relazione conforme}
\eeq
Similar identities hold for any kind of Green functions.\\ 

\subsection{Extendibility beyond the Horizons}

 In order to obtain
 final Wightman functions which are defined also when the two arguments 
are on {\em opposite sides} of a horizon and furthermore,
 Wightman functions which are 
defined on the whole manifold,
 we shall study whether it is possible to {\em analytically}
extend the previous Bertotti-Robinson Wightman functions.
We shall find this possibility holding in the case of $T=1/\be=0$ only.\\
Taking into account the existence of a global
time-like Killing vector $\partial_{T}$,
we expect to find a time-translationally invariant function also with
respect to the  global time $T$, as it happens for the 
Minkowski vacuum in the Rindler wedge theory.\\
Our main idea is to consider the case $\ep=0$, avoiding light-like correlated
arguments, understanding the Wightman functions as proper
functions; furthermore to keep fixed an argument in the 
interior of a certain fixed B-R region and posing the second
near the (future or past) event horizon. Finally we want to translate
the Wightman function from R-N variables into variables $R,T$ (which are
 regular
on the horizon) and to check whether the obtained function of the second
argument is
 {\em analytically} extendible beyond the horizon into an other B-R region.\\
Obviously we have to perform an analogous procedure which starts 
in the latter region and extends the function into the former region.
 It is reasonable to demand that the obtained extended functions are 
the same in both cases.\\

First we consider the  simple case $T=1/\beta = 0$. 
Starting from Eq.~(\ref{wightman zero distribuzione}) and passing to
variables $R$ and $T$ by means of Eq.s~(\ref{r}) and (\ref{t}) in the 
case $\ep=0$ it arises
\beq
W^{\pm}(X,X^{'})_{BR} = 
\frac{(4 \pi^{2})^{-1}\:\:\sin 2R\:\:\sin 2R^{'}}
{(\cos 2R \:-\cos 2R^{'})^{2}+ |\sin 2\vec{R}\: -
\sin 2\vec{R^{'}}|^{2} - 4 \sin^{2}(T-T^{'})}\:,
\label{primo caso}
\eeq
where $sin 2\vec{R}$ means the $3$-vector parallel to $\vec{x}$ carrying
 a length $|\sin 2R|$.\\
This formula holds when both arguments belong to the interior of the {\em same}
 B-R region.
It is now evident that, keeping one point fixed inside a B-R zone
(but  not on the horizon), the above function
is analytic in the second variable, also {\em on the horizon}.
Hence Eq.~(\ref{primo caso}) can be analytically extended
 for arguments inside of two {\em different} B-R zones, too.
 It is important to note that the resulting
function is {\em invariant} under $T-$translations. We also observe that 
the validity of the Haag, Narnhofer and Stein
 scaling prescription \cite{HNS,FH,haag libro,moretti}
 as well as the Hessling
prescription \cite{hessling,moretti} is quite 
straightforward to prove employing the form in Eq.(\ref{primo caso}) of 
Wightman functions;
the same result was proved in \cite{moretti}, but by applying a different 
coordinate frame.\\
We return to the above function later in order to discuss its
interpretation as  distribution after restoring the $\ep$-prescription.\\

Let us consider the case $\be$= finite and look for
possible analytical continuations on the whole manifold.
 We have to translate  the right hand
side of Eq.~(\ref{wightman distribuzione}) into $R$ and $T$ in the
case $\ep=0$
(we shall drop 
the index $BR$ everywhere).\\
We shall analyze separately the different terms which appear therein. \\
First we translate the external  factor
\beq
 F(x,x^{'}):=\frac{\mid \vec{x}\mid
\mid \vec{x^{'}}\mid}{|\vec{x}-\vec{x}^{'}|}\:.
\eeq
when both arguments remain in the same B-R region (for example, in the B-R
 region containing the
 $R$-axis).
 In terms of the global coordinates this reads
\beq
F(X,X^{'})=\left|
\frac{\cos 2T -\: \cos 2R}{\sin 2R} \vec{n}  -
\frac{\cos 2T^{'} -\: \ cos 2R^{'}}{\sin 2R^{'}}\vec{n^{'}}
\right|^{-1}
\:,\label{es}
\eeq
where we used the notation
\beq
\vec{n} := \frac{\vec{x}}{r} \:\:\:\mbox{and} \:\:\: \vec{n}^{'} :=
 \frac{\vec{x}^{'}}{r^{'}}
\eeq
Keeping fixed one argument $X^{'}$
 away from the horizon ($T^{'}=\pm R^{'}$) and considering the function of the
remaining argument $X$,
 one can  demonstrate that there exists
 a region which crosses a part of the horizon where the
 absolute value in  the expression (\ref{es}) does not vanish.
 Eq.~(\ref{es})
defines an analytic function in this region. Furthermore one finds
 the same
 function
starting from the opposite side of the horizon. However, it is important
to point out that the  translational time invariance is lost. 
It is also obvious that 
the {\em coth} in the remaining part of the Wightman function does not 
cancel these ``bad'' terms. 
 Hence, for $T=1/\be>0$,
 it is {\em not} possible to
extend the thermal B-R states to {\em stationary} states
(thermal or not) of the global time $T$.\\
Still choosing  both  arguments in the same B-R region (for example
in the B-R region containing the $R$-axis),
 we analyse the two different  arguments of the {\em coth} in the 
case of $W_{\be}^{+}$ in Eq.(\ref{wightman distribuzione})
\beq
A^{\pm}(x,x^{'}):=[|\vec{x}-\vec{x}^{'}|\pm(t-t^{'})] \:.
\eeq
We shall prove they produce a {\em discontinuity} in the Wightman functions
if we  suppose Eq.(\ref{wightman distribuzione}), translated into Carter-like
coordinates, 
 holds also when the arguments stay on the opposite
side of an horizon.\\  
Stepping over to global null coordinates and rearranging them in
a more useful form we find
\beq
A^{\pm}(X,X^{'})=\frac{1}{2} (\cot U \:+ \cot W )\times \nn
\eeq
\beq
\times
\sqrt{1+\left(\frac{\cot U^{'}\:+\cot W^{'}}{\cot U \:+\cot W} \right)^{2}
 -2 \:\left(\frac{\cot U^{'}\:+\cot W^{'}}{\cot U\:+\cot W}\right)\cos
\theta
} \:+\nn
\eeq
\beq
\pm \frac{1}{2} [ \cot U \:-\cot W - (\cot U^{'}\:- \cot W^{'})]\:,
\label{mostro}
\eeq
where $\theta$ is the angle between $\vec{n}$ and $\vec{n}^{'}$ defined 
above; furthermore,
$U^{'}$, $W^{'}$ and the associated angular coordinates are fixed while
 $U$, $W$
and  the corresponding angular coordinates are varying. In particular we want
to reach the future horizon, $W \rightarrow 0^{+}$.
In this way we find
\beq
\coth A^{-} (W)\rightarrow 1 \:,\nn
\eeq
and
\beq
\coth A^{+} (W)\rightarrow  \coth \left[\cot U\: -
\frac{1+2 \cos \theta}{2}\:\cot U^{'}
+\frac{1-2\cos \theta}{2}\:\cot W^{'}\right] \:.\nn
\eeq
Supposing  Eq.(\ref{mostro})
makes sense also when its arguments are on opposite sides of the future
horizon, we calculate the limit  as the argument $X$ approaches
 the future
 horizon
from the region $T>R$ while the argument $X^{'}$ is fixed in the region $R>T$.
By this way we obtain
\beq
\coth A^{-} (W)\rightarrow -1 \:,\nn
\eeq
and
\beq
\coth A^{+} (W)\rightarrow  \coth \left[\cot U\: -
\frac{1+2 \cos \theta}{2}\:\cot U^{'}
+\frac{1-2\cos \theta}{2}\:\cot W^{'}\right] \:.\nn
\eeq
Thus a {\em discontinuity} appears
which propagates directly into the final form of the function
$W^{+}_\be(X,X^{'})$ because
all the other functions used to build up $W^{+}_\be$ 
are continuous on the horizon,
and in particular $F(X,X^{'})$ is not vanishing there.
Hence, we  cannot suppose the general validity of Eq.~(\ref{mostro})
on the whole manifold {\em sic et simpliciter}. Then, another chance 
is to calculate
a Taylor series (in several variables) of the running argument on the
 horizon, using the limits of the derivatives towards
the horizon, when both arguments stay inside of the same region.
If the convergence radius is not zero this determines an 
 extension of  $W^{+}(X,X^{'})$ beyond the horizon.\\
If we examine the $W$-derivative
we obtain for $W\rightarrow 0^{+}$
\beq
\frac{\partial^{n}\:\:\:\:}{\partial W^{n}}\coth A^{-} (W)\rightarrow 0 \:,\nn
\eeq
and
\beq
\frac{\partial^{n}\:\:\:\:}{\partial W^{n}}\coth A^{+} (W)\rightarrow
\mbox{finite expression} \nn
\eeq
It arises from the result of the former limit that the convergence radius of
 the Taylor series
 of the function $A^{-}(X,X^{'})$ ($X^{'}$ fixed) vanishes on the horizon
and thus it is not possible to reconstruct the  function on {\em both sides}
 of the horizon with the help of just this Taylor series.  
 The function does not admit any
analytical extension beyond the horizon. It is simple to conclude that
also 
the function $W^{+}_\be(X,X^{'})$ ($\be<+\infty$)
 can not be analytically extended beyond the horizon analytically.\\
Here, it is important to remind that the B-R KMS
states with $\be>0$
(as well as the B-R vacuum state at $T=1/\be=0$) satisfy
the HNS prescription also  on the horizon \cite{moretti}, but
(differently from the vacuum state) 
they carry an infinite  renormalized
stress tensor on the horizon \cite{AHL} and they do {\em not}
 satisfy
 Hessling's prescription \cite{moretti,moretti2}.\\

\subsection{A Bisognano-Wichmann-like Theorem}

Now we return to the case $\be=0$.
 We shall prove a {\em Bisognano-Wichmann-like theorem}
 as our third result.\\
We interpret the Wightman functions defined in
Eq.~(\ref{wightman zero distribuzione}) as distributions working
on four-smeared functions \cite{fulling,haag libro,kaywald,kay,higuchi}
with support enclosed in the B-R considered region. 
 It is  possible to prove
 that  these
 Wightman functions coincide with the
Wightman functions of the vacuum state defined by quantizing with respect 
to the global coordinates $R$ and
$T$  when we restrict the latter in the  interior of a R-N sub-manifolds.\\
By the {\em GNS theorem} or similar theorems \cite{haag libro,kaywald} we
 are able to
 extend this property from the Wightman functions onto the respective
  quantum states.  
 This fact  corresponds to  the
  Bisognano-Wichmann theorem in Minkowski space-time \cite{sewell}. 
In this sense the analog to the Unruh-Rindler temperature in the
 ``B-R wedges'' is exactly 
$T=1/\be=0$ and thus the KMS conditions does not appear, but
the dependence of $t-t^{'}$ remains in the Wightman functions of the
analog to  the $\be=2\pi$-Rindler-KMS state. The 
$\be=2\pi$-Rindler-KMS state corresponds to  the
B-R vacuum now and  the Minkowski vacuum is represented 
by the Carter-like global vacuum.  \\

We shall prove our theorem employing the following way.
First, we shall express  Wightman functions in therms of 
{\em Feynman propagators}, then, we shall prove that 
the coincidence inside of B-R submanifolds
of the Feynman propagators involves
 the coincidence of Wightman functions there.
Finally, we shall prove the coincidence of Feynman propagators.\\
We can extract the  Wightman functions 
from the Feynman propagator using
well-known properties working in {\em static, globally hyperbolic}
 space-times \cite{fulling,haag libro}. In the case of the 
B-R space-time and also in the case of our complete Carter-like manifold, 
the following
identities arise directly from the analog identities 
which hold in the respective
conformal related ultrastatic manifold, using Eq.s~(\ref{identita'})
and (\ref{esu}).\\
Let us start with the first part of the proof. 
In general coordinates
\beq
iG_F =\theta(\tau-\tau^{'})\:W^+ +\theta(\tau^{'}-\tau)\:W^-
=\theta(\tau-\tau^{'})\:W^+ + \theta(\tau^{'}-\tau)\:W^{+\ast}\:, \nn
\eeq
 where $G_F$ denotes the Feynman propagator.
It arises from the above identity 
\beq
i\theta(\tau-\tau^{'})G_{F}=\theta(\tau-\tau^{'})W^{+}\:\:\:\: \mbox{and}
\:\:\:\:i\theta(-\tau+\tau^{'})G_{F}=
\theta(-\tau+\tau^{'})W^{+\ast}\:,\label{soprat}
\eeq
and thus:
\beq
W^{\pm}=i\theta(\pm(\tau-\tau^{'}))G_{F}-i 
\theta(\pm(\tau^{'}-\tau))G_{F}^{\ast}\:. 
\label{confronto}
\eeq
Suppose now the coincidence of Carter-like propagator 
and Bertotti-Robinson propagator were proved 
 inside of a B-R sub manifold, then, the coincidence
of Wightman functions follows as well. In fact, whenever the arguments of 
the Wightman functions
are {\em space-like} related, the field operators commute
 and thus $W^+\equiv W^-\equiv G_F$ from the previous 
formulas. Then, the coincidence of Wightman functions 
follows from the coincidence of Feynman propagators.
On the other hand, if the arguments of the Wightman functions are 
{\em time-like} or {\em light-like} related,
 the functions $\theta(T-T^{'})$ and 
$\theta(t-t^{'})$ which
appear in Eq.(\ref{confronto}) as well as in the Feynman propagators 
 trivially coincide  and thus the Wightman functions coincide, too.\\
We have to prove the coincidence of 
Feynman propagators in the remaining of this section.\\
The Feynman propagator   of
 a massless field in the Minkowski space-time is well-known (see
for example \cite{itzykson}).
 Taking into account Eq.~(\ref{identita'})
 we get
\beq
G_{F}(x,x^{'})_{BR}= 
\frac{-i}{4\pi^{2}}\frac{rr^{'}}{|\vec{x}-\vec{x}^{'}|^{2} -(t-t^{'})^{2}}
- \frac{ rr^{'}}{4\pi} \delta(|\vec{x}-\vec{x}^{'}|^{2}-(t-t^{'})^{2}) \:. 
\label{feynmanBR}
\eeq
We introduce  the Feynman propagator in  Carter-like manifold.
This can be calculated from  Feynman propagator in the  
Einstein's static
universe with spatial radius $\rho=1$ (which is our case) 
 for a   conformally 
coupled scalar field.  We report this  in {\bf Appendix C}.
\beq
G_{F}(T-T^{'}, \vec{R}, \vec{R}^{'})_{CEU}= \nn
\eeq
\beq
\frac{i\:\sin 2R \: \sin2R^{'}}{4\pi^{2}}
\frac{1}{2-2\cos\sigma -4 \sin^{2}(T-T^{'})}  + \nn
\eeq
\beq
+ \frac{\sin 2R \sin 2R^{'} }{4\pi}
 \sum_{n\in \Z} \frac{\sigma + 2\pi n}{\sin \sigma}
\delta(  (2T-2T^{'})^{2}-(\sigma+2n \pi)^{2} ) 
\:,\label{feynmanCEU}
\eeq
where $\sigma$ is the {\em minimal} geodesical length between the points
determined by $\vec R$ and $\vec{R}'$ on a $3-$sphere $S^3$. 
Using our coordinates $\vec{X}\equiv (R, \theta, \varphi)$
 on the above $3$-sphere, it is possible to prove that $\sigma$ satisfies
\beq
2-2\cos\sigma(\vec{X},\vec{X})=(\cos 2R\: -\cos 2R^{'})^{2} + |\sin \vec{2R}
\: -\sin \vec{2R}^{'}|^{2}\:.\label{sigma}
\eeq
Now we prove that, {\em in the interior a B-R submanifold}, 
the Feynman propagator previously evaluated
 coincides with the Feynman propagator in Eq.~(\ref{feynmanBR}). 
In order to prove this coincidence in a B-R zone,  it is sufficient to
demonstrate the following identity 
\beq
\sin2R \:\sin 2R^{'}
\sum_{n\in \Z}\frac{1}{4\pi}\frac{\sigma+2\pi n}{\sin\sigma}
\delta((2T-2T^{'})^{2}-(\sigma +2n\pi)^{2} ) =  \nn
\eeq
\beq
= \frac{ rr^{'}}{4\pi} \delta(|\vec{x}-
\vec{x}^{'}|^{2}-(t-t^{'})^{2}) \:.\label{ultimissima}
\eeq
In fact, the first term on the right hand side of Eq.~(\ref{feynmanCEU})
  trivially coincides with the first term on the right hand side of 
Eq.~(\ref{feynmanBR})
if one uses Eq.~(\ref{sigma}). This is nothing but Eq.(\ref{primo caso}).\\
Let us prove  identity (\ref{ultimissima}),
 reminding that both arguments
belong to the interior of a R-N zone and noting that the minimal
geodesical length $\sigma$ is contained in the interval
$[0, \pi]$ and thus $\sin \sigma = |\sin
\sigma |$.
\beq
   \sin2R \:\sin 2R^{'}
\sum_{n\in \Z}\frac{1}{4\pi}\frac{\sigma+2\pi n}{\sin\sigma}
\delta((2T-2T^{'})^{2}-(\sigma +2n\pi)^{2} )  = \nn
\eeq
\beq
\sum_{n\in \Z}
\frac{ \sin 2R \:\sin 2R^{'} (\sigma+ 2\pi n)}{4\pi \sin (\sigma+2\pi n)}
\left[ \frac{\delta(\sigma+2\pi n-(2T-2T^{'}))}{2(\sigma+2\pi n)}+
\frac{\delta(\sigma+2\pi n +(2T-2T^{'}))}{2(\sigma+2\pi n)} \right]=\nn
\eeq
\beq
=\frac{\sin 2R\: \sin 2R^{'}}{4\pi} \delta(-2 \cos\sigma +
 2\cos (2T-2T^{'}) ) =\nn
\eeq
\beq
=\frac{ \sin 2R\: \sin 2R^{'}}{4\pi} \delta \left(
\frac{-2 \cos\sigma +
 2\cos (2T-2T^{'})}{\sin 2R \: \sin 2R^{'}}\sin 2R \: \sin 2R^{'} \right)= \nn
\eeq
\beq
=\frac{ \sin 2R\: \sin 2R^{'}}{4\pi} \delta\left(
\frac{|\vec{x}-\vec{x}^{'}|^{2}-(t-t^{'})^{2}}{r r^{'}}
\sin 2R \: \sin 2R^{'} \right)\:. \nn
\eeq
We used Eq.(\ref{primo caso}) 
(holding  inside of any B-R region) once again
in the argument of the  delta function.\\
Considering $t$ as the integration variable and keeping $r$, $t^{'}$, $r^{'}$ 
 fixed,
using standard manipulations on delta function,
 we find  that the above term can also be written as
\beq
\frac{1}{4\pi} rr^{'} \delta(|\vec{x}-\vec{x}^{'}|^{2}-(t-t^{'})^{2})\:.\nn
\eeq 
We just 
obtained the second term on the right hand side of Eq.~(\ref{feynmanBR}), i.e.,
we proved the coincidence of $G_{F\:BR}$ and $G_{F\:CEU}$ in the interior of
a B-R zone. \\

Just two technical notes  to conclude.\\
First, we   write 
Wightman functions of the Carter-like manifold 
in a more concise form.
Using the identity
\beq
\frac{1}{x\pm i\ep}= Pv\: \frac{1}{x} \mp i \pi \delta(x) \nn
\eeq
where $Pv$ denotes the {\em principal value}
and  taking into account that the first term on the right hand side of 
Eq.~(\ref{feynmanCEU}) is to be understood just in the sense of 
the  principal value
 \cite{fulling}, it arises from Eq.~(\ref{confronto})
\beq
W^{\pm}(T-T^{'}, \vec{R}, \vec{R}^{'})_{CEU}= \nn
\eeq
\beq
\frac{\sin 2R \: \sin2R^{'}}{4\pi^{2}}
\frac{1}{2-2\cos\sigma -4 \sin^{2}(T-T^{'}\mp i \ep)} \:. \label{nota}
\eeq  
Finally, we can also observe that
the coincidence of the Wightman functions (in a B-R zone) 
in the case of $\ep=0$ is equivalent 
to the coincidence of the {\em Hadamard functions} therein. 
We can  calculate the Hadamard functions as
\beq
G^{(1)} := W^{+}+W^{-}\:.\nn
\eeq 
In the case of the B-R metric, the Hadamard function reads (to be understood 
in the sense of the principal value)
\beq
G^{(1)}(x,x^{'})_{BR}= \frac{1}{2\pi^{2}}
\frac{rr^{'}}{|\vec{x}-\vec{x}^{'}|^{2}-(t-t^{'})^{2}} \:, \label{hadamardBR}
\eeq
on the other hand, in the case of the global metric, the Hadamard function 
reads 
 \beq
G^{(1)}(X,X^{'}) = \nn
\eeq
\beq
=\frac{\sin 2R \: \sin2R^{'}}{2\pi^{2}}
\frac{1}{2-2\cos\sigma -4 \sin^{2}(T-T^{'})} \:. \label{hadamardCEU}
\eeq
These functions coincide as proved above in Eq.(\ref{primo caso}).

\section{Conclusions and  Outlooks on Exact Extremal R-N Black Holes}

 The most important result of this paper is the proof of the 
coincidence of the 
global  Carter-like  vacuum and the Bertotti-Robinson vacuum.
 Notice that the global vacuum
which is represented by a {\em pure} state also inside of a B-R submanifold,
 has {\em vanishing entropy} there, despite of the presence of horizons.
This is probably due to the fact that the horizons do not separate
different spatial regions differently from the Minkowski-Rindler and
Kruskal-Schwarzshild case. In these latter cases the whole spatial Cauchy 
surface at $t=0$ (where $t$ is the Minkowski or Kruskal time) is separated
into two Cauchy surfaces within two (Rindler or Schwarzschild) wedges.
Let us consider the Minkowkian case. 
Formally employing a von Neumann approach,
 this separation of the  Minkowski Cauchy surface involves
 a ``separation'' of the field Hilbert space  
 which results to be a tensorial product of two Hilbert spaces
related with the two Rindler wedges. Then, a pure state (with 
vanishing entropy) 
of the whole
Hilbert space appears as a mixed state (with a non vanishing entropy) 
in each  factor  Hilbert space. However, in our case the situation is 
more complicated due to the fact that the $T=0$ surface is not a Cauchy
surface. \\     
Another point is the following.
Supposing that physically sensible KMS (including the case $T=1/\be=0$)
quantum states are analytically 
extendible on the whole manifold, 
we have to accept {\em only} $T=0$ as possible temperature without the use
of further physical requests. 
 This fact arises regardless of all the  topological consideration
on {\em conical singularities} in the Euclidean formulation. In fact, 
our manifold does not produce conical singularities for any
choice of the Euclidean time period $\be$ and thus, in the framework
of the Euclidean formalism, one should accept every value for the 
temperature to be possible. \\
We expect that it should be  possible to develop further the analogy
between Minkowski space-time and our model in order to prove the above 
coincidence of vacuum states also for the case of the extremal 
Reissner-Nordstr\"{o}m space-time and the Carter space-time.
In the case of the extremal R-N black hole, the hardest
 problem is to deal with the {\em time-like} singularity in the region
beyond the horizon. It  is not possible to develop a standard quantum
field theory there. However it seems to be possible to employ a more 
general theory based on the Kay F-locality \cite{kay,higuchi} (or 
something similar) inside the manifold
resulting  from  Carter's manifold  by
excluding all the points belonging to the time-like singularity.
Following this way, it should be possible 
to define a global {\em advanced-minus-retarded} fundamental solution  
which agrees to that one defined inside of each B-R 
manifold.\footnote{Remind that this ``Green function'' does not depend on
the considered quantum state.}
Using Carter's coordinate, the idea is to analytically extend beyond the 
horizons  
the (thermal) Hadamard  function built up inside of a B-R region, 
  defining a global Wightman function and thus a
global quantum state.\\
We expect that {\em only} the B-R vacuum defines a similar global extension.   
Furthermore, if this is  proved to be correct, following the results
in \cite{cvz},  no quantum  one loop
 corrections (generally singular) which arise from 
the  (massless scalar)
 fields propagating
outside of the extremal R-N black hole, need to be added 
to the  gravitational entropy.

\ack{We would like to thank Luciano Vanzo for his bright lectures on several 
topics of this paper as well as  Marco Toller,  
Sergio Zerbini and Giuseppe Nardelli for many helpful hints.\\
 Stefan Steidl would like to thank the Dipartimento di Fisica 
dell'Universit{\`a} di Trento for its kind hospitality during his stay in 
Trento and especially the persons mentioned above, including also Guido 
Cognola, for the cordial atmosphere they provided. }

\section*{Appendix A. Approximations near the Horizons in Carter's Map}

Let us  consider $ds^{2}$ in Eq.(\ref{RNC}) as the  quadratic form
\beq
ds_{x}^{2}(\vec{X}):= g_{\mu\nu}(x)\:dX^{\mu}\:dX^{\nu}\:, \nn
\eeq
where $dX^{\mu}\equiv (dT,dR,d\theta,d\varphi)$ are the components
of the $4$-vector $\vec{X}$ at $x\equiv (T,R,\theta,\varphi)$.\\
In order to deal with the approximated metric near different
points on the horizon,
we shall consider the following expansion as $\bar{r}
\rightarrow 1$ (i.e. near the horizons) which arises
{from} the definition of $r^{\ast}$ (Eq.(\ref{ar}))
\beq
\left( 1-\frac{1}{\bar{r}}\right)^{2} = \frac{1}{r^{\ast 2}} (1+ O ((\bar{r}-1)
\ln |\bar{r}-1|)) \:, \label{sviluppo base}
\eeq
and also, trivially
\beq
\bar{r}= 1+ (\bar{r}-1) = 1+ O(\bar{r}-1)\:. \label{banale}
\eeq
Let us define the approximated form of the above metric 
\beq
ds_{0x}^{2}(\vec{X}):= \frac{1}{r^{\ast 2}} \csc^{2}(R+T)\csc^{2}(R-T)\:
(-dT^{2}+ dR^{2}) + d\Om_{2}(\vec{X})\:, \label{dszero0}
\eeq
where we posed also
$d\Om_{2}(\vec{X}) := d\theta^{2}+\sin^{2}\theta\: d\varphi^{2}$.\\
Thus, it holds by definition
\beq
ds_{x}^{2}(\vec{X}) = ds_{0x}^{2}(\vec{X}) + (ds_{0x}^{2}(\vec{X}) -
d\Om_{2}(\vec{X})) O_{\vec{X}} ((\bar{r}-1)
\ln |\bar{r}-1|)) + O_{\vec{X}}(\bar{r}-1) d\Om_{2}(\vec{X}) \:. \nn
\eeq
Taking the  leading order only as $\bar{r} \rightarrow
1$ we have
\beq
ds_{x}^{2}(\vec{X}) \sim ds_{0x}^{2}(\vec{X})\:. \nn
\eeq
The metric $ds^2_0$ can be written in a more useful form by employing
 the formula
\beq
\frac{1}{r^{\ast 2}} \csc^{2}(R+T)\csc^{2}(R-T) = \frac{1}{\sin^{2} 2R}
\label{2R}\:;
\eeq
in this way  we find the static
metric of Eq.(\ref{dso})
\beq
ds_{0x}^{2}(\vec{X}):= \frac{4}{\sin^{2} 2R}\:
(-dT^{2}+ dR^{2}) + d\Om_{2} \label{dszero}\:
\eeq
The vector field $\partial_{T}$ defines an {\em approximated}
Killing vector inside the regions
where $ds^{2}_{0}$ approximates to  $ds^{2}$.\\
{From} now on, we shall drop the index $x$ and the explicit dependence
from $\vec{X}$ for sake of simplicity.\\

Let us specify the regions, in Carter's picture, where we may employ
the previous approximated
form of the metric using Carter's coordinates as well as R-N coordinates.\\
We start by  considering  the part of the future horizon between
the origin $O$
and its opposite point $O^{'}$ (see {\bf figure 2}).
We define the coordinate $W$ and $U$ in the two regions
\begin{eqnarray}
R-T &=:& W \sim 0 \:\:\:\:\: R>T\:\:\: or \:\:\: R<T \label{prima}\\
R+T &=:& U \:\:\: \mbox{finite}\:. \label{seconda}
\end{eqnarray}
By using Eq.~(\ref{r estesa}) one finds that, fixing $\ep>0$,
$\bar{r}\rightarrow 1^{\pm}$ {\em uniformly} in $U
\in [ \ep, \pi/\sqrt{2}\:-\ep]$
as  $W \rightarrow 0^{\pm} $.
We conclude that  it is possible to use the
form of the metric of Eq.s~(\ref{dszero0}) and ({\ref{dszero}).\\
Let us consider the form of the metric in the considered region 
employing R-N coordinates. We shall find the {\em Bertotti-Robinson} metric.
We start by consider that $ds^{2}$  reads in terms of $U$ and $W$
Carter's null
coordinates
\beq
ds^{2} \sim \frac{1}{r^{\ast 2} \sin^{2}W \:\sin^{2} U} dU dW + d\Omega_{2}
\label{settima}\:.
\eeq
 Employing 
the following identities
\begin{eqnarray}
du &=& \frac{1}{\sin^{2} U} dU \nn \:,\\
dw &=& \frac{1}{\sin^{2} W}    dW \nn
\end{eqnarray}
and 
\beq
du \: dw = \frac{1}{\sin^{2} W \: \sin^{2} U } dU dW \:; \nn
\eeq
and translating  into R-N null coordinates, we  finally find
\beq
ds^{2} \sim \frac{1}{r^{\ast 2}} du\: dw + d\Omega_{2} \label{settimaprimo}\:.
\eeq
Thus, in coordinates $r^{\ast}$, $t$, the metric of Eq.(\ref{settima})
reads \cite{referee,AHL,moretti}
\beq
ds^{2} \sim \frac{-dt^{2} + dr^{\ast 2}}{r^{\ast 2}} + d\Omega_{2}
= \frac{-dt^{2} + dr^{2} + r^{2} d\Omega_{2}}{r^{2}} \label{ottava} \:,
\eeq
where we posed $r:= -r^{\ast}$ if $R>T$, or $r:=r^{\ast}$ if $R<T$.\\
This metric is the well-known
{\em Bertotti-Robinson metric} \cite{BR}.\\

Now, let us  consider the regions near the extremal points of the horizon.
We start by considering the ``origin''
$O \equiv (R=0, T=0)$\footnote{Really $O$ is a
{\em 2-sphere}.}
\begin{eqnarray}
T&\sim& 0 \:\:, \:\: T > 0 \:\:,\nonumber\\
R&\sim& 0 \:\:, \:\: R > 0 \:\:. \label{regione}
\end{eqnarray}
We observe that $r^{\ast}$ and $\bar{r}$ are defined in terms of
$R$ and $T$. $\bar{r}$ reaches {\em uniformly} its value $1$ in the sense of
$\R^{2}$ when $(R,T) \rightarrow (0,0) $ in any
wedge of the form ($\ep>0$)
\beq
R > (1-\ep)\mid T \mid\:, \:\:\:\: T>0\:, \label{wedge}
\eeq
or, inside the region beyond the future horizon
\beq
\frac{R}{\ep}> T > (1+\ep) R \:, \:\:\:\: R>0 \:\:\:\: T>0 \:. \label{wedge2}
\eeq
Thus we can use Eq.~(\ref{dszero}) for the approximated metric.\\
We can notice another interesting fact. 
By means of Eq.~(\ref{2R}), we also obtain in these regions
\beq
Q  = \frac{1}{R^{2}}+ O(R,T) \nonumber \:,
\eeq
where $O(R,T)$ is an  infinitesimal function as $(R,T) \rightarrow (0,0)$.
Thus, the metric inside of the considered wedges,
employing the leading order approximation  as $(R,T) \rightarrow (0,0)$,
reads
\beq
ds^{2} \sim \frac{1}{R^{2}}
\left( -d T^{2}+d R^{2}\right) + 1\:d\Omega_{2} \nn \:,
\eeq
or
\beq
ds^{2} \sim \frac{1}{R^{2}}
\left( -d T^{2}+d R^{2} + R^{2}\:d\Omega_{2} \right) \label{brc} \:.
\eeq
We found
 the Bertotti-Robinson metric also in Carter's coordinates.\\
It can easily be proved  by hand that the above approximation also holds
in the R-N region, dropping the constraint $T \neq 0$.
Furthermore, due to the symmetry of the manifold, similar
calculations can be performed for
 $T<0$. Thus the B-R metric, in the limit of
``little'' $R$ and $T$, holds for all the
wedges of the form
\beq
R > (1-\ep)\mid T \mid\:, \:\:\:\: R>0 \label{wedge general}
\eeq
and
\beq
\frac{R}{\ep}> \mid T \mid > (1+\ep) R \:, \:\:\:\: R>0 \:.
\label{wedge general2}
\eeq
Let us consider the form of the metric in R-N coordinates 
in the region defined by Eq.~(\ref{regione}).\\
 Keeping the divergent leading
order in $R$ and $T$ in  Eq.~(\ref{regione}),
 Eq.(\ref{r estesa}) reads
\beq
r^{\ast} = -\frac{1}{2(T+R)}+\frac{1}{2(T-R)} + O(R,T) \sim
\frac{R}{T^{2} -R^{2}}\:\:\:\:\:\: ( R>0 )\label{r approx}\:;
\eeq
for the coordinate $t$ we obtain similarly
\beq
t = -\frac{1}{2(T+R)}-\frac{1}{2(T-R)} + O(R,T) \sim
-\frac{T}{T^{2} -R^{2}}
\:\:\:\:\:\: ( T >0 )      \label{t approx}\:.
\eeq
It follow from these
\beq
R^{2}-T^{2} \sim (r^{\ast 2}-t^{2})^{-1}
\eeq
Using the  latter three equations, we can write 
$R$ and $T$ in Eq.(\ref{brc}) in terms of $t$ and $r:= -r^{\ast}$
in the  R-N region, or $r:= r^{\ast}$ beyond the horizon. Thus,  we recover
the approximated form  of the
metric also in R-N coordinates in the respective
regions.
In fact, we get the (dominant  order approximated)
inverse relations of Eq.s~(\ref{r approx}) and (\ref{t approx})
\begin{eqnarray}
R&\sim&\frac{r^{\ast}}{t^{2}-r^{\ast 2}}\\
T&\sim&-\frac{t}{t^{2}-r^{\ast 2}}\:.
\end{eqnarray}
Substituting these results in  Eq.~(\ref{brc}) we find
the Bertotti-Robinson metric once again 
\beq
ds^{2} \sim \frac{-dt^{2} + dr^{2} + r^{2} d\Omega_{2}}{r^{2}}\:.
\eeq
Let us examine  the metric near the  ``point'' at infinity
(really a {\em 2-sphere}):\\
$O^{'}\equiv (T=\frac{\pi}{2},
R=\frac{\pi}{2})$.\\
We shall just sketch the approximation because this is very similar to
the previous one.\\
Starting from the ansatz
\begin{eqnarray}
T &=&\frac{\pi}{2}-\hat{T}\:, \\
R &=&\frac{\pi}{2}\pm \hat{R}\:,
\end{eqnarray}
which implies $dT^{2}=d\hat{T}^{2}$ and $dR^{2}=d\hat{R}^{2}$
and considering the limit $ (\hat{R},\hat{T})\rightarrow (0,0)$
as in the previous case, we find
\beq
Q\sim \frac{1}{\hat{R}^{2}}\:,
\eeq
whatever the sign in front of $\hat{R}$ may be.\\
For $r^{\ast}$ we obtain the formula
\beq
r^{\ast} \sim \frac{ \pm \hat{R}}{\left( \hat{T}^{2} -
\hat{R}^{2}\right)}\:.
\eeq
We see that, in order to restore the Bertotti-Robinson metric, the only
possible choice for the sign in front of $\hat{R}$ is $-$.
In fact, 
this guarantees that $r^{\ast}$ tends to $-\infty$ (i.e.
$\bar{r}\rightarrow 1^{+}$) as $ (\hat{R},\hat{T})\rightarrow (0,0)$
and $\hat{T}>\hat{R}$
( coming from the interior of the R-N region ). 
On the other hand this also guarantees that
$r^{\ast}\rightarrow +\infty$ (i.e. $\bar{r}\rightarrow 1^{-}$) when
$\hat{T}<\hat{R}$, which is when the horizon is approached
from outside of the R-N region.\\
Therefore, if we change coordinates $T\rightarrow \hat{T}+\frac{\pi}{2}$
and $R\rightarrow \hat{R}+\frac{\pi}{2}$ we find  the Bertotti-Robinson metric
in terms of $\hat{T}\ll 1$ and $\hat{R}\ll 1$ within wedges of $\hat{T}$
and $\hat{R}$ similar to those previously found.\\
It can easily be proved that by translating the obtained metric into R-N
coordinates ${t,r,\theta,\varphi}$ and using  usual approximations,
the metric results to be  the Bertotti-Robinson metric
 as in the previous case.\\
In the first case we used the  null coordinates $U$, $W$ and $u$, $w$,
respectively, instead of
the usually employed
space-like and time-like ones used near $O$ and $O^{'}$.
However we can point out
how the first case formally includes the remnant ones
if we do the limit
$U \rightarrow 0^{+}$ or $ U \rightarrow \pi^{-}$ in
Eq.~(\ref{settima}) and translate the result into the variables $R$ and $T$.\\
Furthermore we studied the manifold near a particular
future horizon, but obviously, due to the evident
symmetries of Carter's  manifold,
we may repeat all the previous calculations for all the event
horizons (past or future) therein.\\

Finally, let us   
consider the form in Eq.~(1) of the metric, i.e., the metric 
directly expressed in R-N coordinates
in the R-N region and in the region containing the irremovable
singularity.\\
It is easy to prove that, keeping $t\in \R$ fixed,
the metric transforms into the B-R metric as 
$\bar{r} \rightarrow 1^{\pm}$ ( or $r \rightarrow +\infty$ ).\\ 
Looking at {\bf figure~2} one recovers this path to approach the horizon
as falling into the limit point $O$ for $\bar{r}>1$ or into the limit point
$O^{'}$ for $\bar{r}<1$.
Still looking at {\bf figure~2}, we see that, 
in order to reach the remaining points of the horizon also the variable 
$t$ must be increased or decreased, respectively, towards $\pm\infty$. 
Using the R-N picture, these paths approach
the vertical line $\bar{r}=1$ but ``cross'' it only at infinity (in time).

\section*{Appendix B. Approximated Wightman Functions }

In {\bf section 2} we supposed the Bertotti-Robinson Wightman functions
 approximate to the Reissner-Nordstr\"om Wightman functions 
 near the horizons because the Bertotti-Robinson metric approximates to the
 Reissner-Nordstr\"om metric there.
 However, the normalization of normal modes
usually used defining the Wightman functions  depends on the integration over
the
whole spatial manifold and not only on the region near the horizon.
Thus, our hypothesis requires further explanations. \\
 We shall prove that it is possible to
overcome this problem, at least formally, dealing with
{\em static} metrics and KMS states.
 In fact in this case one recovers by the KMS
condition
\cite{HNS} ($x \equiv (\tau, \vec{x})$)
\beq
< \phi(x_{1}) \phi(x_{2}) >_{\beta}\: = \frac{i}{2\pi}
\int_{-\infty}^{+\infty} G(\tau_{1}+\tau,
\vec{x}_{1} \mid \tau_{2}, \vec{x}_{2})
\frac{e^{\be \om}}{e^{\be \om}-1} e^{i\om \tau}\: d\tau d\om
\label{unohaag} \:,
\eeq
where the distribution $G$ is the {\em commutator} of the fields.
This distribution is {\em uniquely} determined \cite{HNS} by the fact that
it is a solution  of the Klein-Gordon equation in both arguments,
vanishes for equal times $\tau_{1} = \tau_{2}$ and is normalized by
the ``local'' condition
\beq
g^{\tau\tau}\sqrt{-g}\frac{\partial}{\partial \tau_{1}}
G(x_{1}, x_{2}) \mid_{\tau_{1}=\tau_{2}}
= \delta^{3}(\vec{x}_{1}, \vec{x}_{2}) \label{duehaag}
\eeq
The above $3$-delta function is usually understood as
\begin{eqnarray}
\delta^{3}(\vec{x}_{1}, \vec{x}_{2})&=&0 \:\:\:\: \mbox{for } \:\:\:\:
\vec{x}_{1} \neq \vec{x}_{2} \nn\\
\int \delta^{3}(\vec{x}_{1}, \vec{x}_{2})\:\: d\vec{x}_{2} &=& 1 \nn
\end{eqnarray}
By the previous, spatially ``local''
formulas  we expect  that
the function $G$ calculated with
the ``true'' static metric becomes the function $G$ calculated by using
the approximated static  form of the metric inside of a certain
static region
$\delta\Sigma \times \R$ (where  $ \tau \in \R$) as $\delta\Sigma$ shrinks
around a $3$-point. Considering $(\delta\Sigma,\tau_{0})$
as a {\em Cauchy surface},
the above result should come out  inside of the ``diamond-shaped'' four-region
causally determined by $(\delta\Sigma,\tau_{0})$  at least. But, studying
the form of the light cones near  event horizons of the form
$|\vec{x}|=r_{0}$ and  $\tau \in \R$,
it can be simply  proved
that this four-region will tend to
contain  the whole $\tau-$axis if $\delta\Sigma$
approaches the event horizons.\\
In the same way, using Eq.~(\ref{unohaag}), we could expect such a
property for thermal Wightman functions, too. The case
of zero temperature, regarded as the limit
$\be \rightarrow +\infty$, is included.\\
In the case of an extremal R-N black hole, the Bertotti-Robinson
metric approximates to the R-N metric along any horizon, for $\bar{r}>1$
and  $\bar{r}<1$ at any time $t\in \R$. This fact simply
follows from the discussion
in {\bf Section 1}.

\section*{Appendix C. Feynman Propagator in the Carter-like Manifold}

Let us start from the Feynman propagator of a scalar field
propagating in the  Einstein's static universe. We
shall prove Eq.(\ref{feynmanCEU}) for the Feynman propagator on the
conformally related Carter-like manifold using Eq.(\ref{esu}).\\
Camporesi \cite{camporesi},
employing heat kernel methods\footnote{Remind that our 
coordinates are  not those which are
 usually used
to describe $S^{2}$ also Einstein's static universe
 as  they contain a factor $2$.},
obtained  the following Feynman propagator 
for a scalar {\em conformally coupled} field
\beq
G_{F}(T-T',\sigma)_{m}= \nn
\eeq
\beq
\frac{im}{8\pi}\sum_{n\in \Z}\frac{\sigma + 2\pi n}{\sin \sigma}
\frac{H^{(2)}_{1}(m[(2T-2T^{'})^{2}-(\sigma + 2n \pi)^{2}-i\ep]^{1/2})}
{[i\ep-(2T-2T^{'})^{2}+(\sigma + 2n \pi)^{2}]^{1/2}}\:, \label{feyman}
\eeq
where $m$ is the mass of the field, $\sigma$ is the  {\em minimal}
geodesical length between two points on
$S^{3}$ and $H^{(2)}_{1}$ is a Hankel function of the second kind of order
$1$.
Furthermore,
using our coordinates $\vec{X}\equiv (R, \theta, \varphi)$
 on the above $3$-sphere, it is possible to prove that $\sigma$ satisfies
\beq
2-2\cos\sigma(\vec{X},\vec{X})=(\cos 2R\: -\cos 2R^{'})^{2} + |\sin \vec{2R}
\: -\sin \vec{2R}^{'}|^{2}\:.\nn
\eeq
Let us to consider the massless case as the limit
 $m \rightarrow 0^{+}$ in Eq.~(\ref{feyman}). We find
\beq
G_{F}(T-T',\sigma)= \nn
\eeq
\beq
-\frac{1}{4\pi} \sum_{n\in \Z} \frac{\sigma + 2\pi n}{\sin \sigma}
 \left\{  \frac{i}{ \pi [ (2T-2T^{'})^{2}-(\sigma + 2n\pi)^{2} ] } -
\delta(  (2T-2T^{'})^{2}-(\sigma+2n \pi)^{2} )  \right\}\:.\nn
\eeq
We can explicitly carry out the summation over the terms which
do not contain delta functions obtaining
\beq
-\frac{1}{4\pi} \sum_{n\in \Z} \frac{\sigma + 2\pi n}{\sin \sigma}
  \frac{i}{\pi [ (2T-2T^{'})^{2}-(\sigma + 2n\pi)^{2} ]} =\nn
\eeq
\beq
 =\frac{-1}{4\pi^{2}\sin\sigma}\:\:\frac{i}{4}
\left\{ \cot\left[ \frac{2T-2T^{'}-\sigma }{2} \right] -
\cot\left[ \frac{2T-2T^{'}+\sigma }{2} \right]  \right\} = \nn
\eeq
\beq
 = \frac{-i}{16 \pi^{2}} \csc\frac{2T-2T^{'}-\sigma}{2}\:
\csc\frac{2T-2T^{'}+\sigma}{2} = \frac{i}{4\pi^{2}}
\frac{1}{2\cos (2T-2T^{'})-2 \cos \sigma} = \nn
\eeq
\beq
 = \frac{i}{4\pi^{2}}
\frac{1}{2-2\cos\sigma -4 \sin^{2}(T-T^{'})}\:.\nn
\eeq
Finally, using Eq.(\ref{esu}), we may prove Eq.(\ref{feynmanCEU})
\beq
G_{F}(T-T^{'}, \vec{R}, \vec{R}^{'})_{CEU}= \nn
\eeq
\beq
\frac{i\:\sin 2R \: \sin2R^{'}}{4\pi^{2}}
\frac{1}{2-2\cos\sigma -4 \sin^{2}(T-T^{'})}  + \nn
\eeq
\beq
+ \frac{\sin 2R \sin 2R^{'} }{4\pi}
 \sum_{n\in \Z} \frac{\sigma + 2\pi n}{\sin \sigma}
\delta(  (2T-2T^{'})^{2}-(\sigma+2n \pi)^{2} )
\:. \nn
\eeq

\end{document}